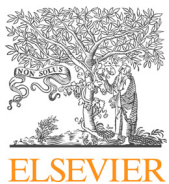
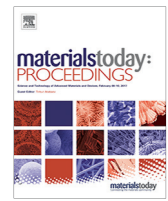

# Effect of the BaO-Na$_2$O-Nb$_2$O$_5$-P$_2$O$_5$ glass addition on microstructure and dielectric properties of BNN ceramics

A. Ihyadn [a,*], D. Mezzane [a,b], M. Amjoud [a], A. Lahmar [b], L. Bih [d], A. Alimoussa [a], I.A. Luk'yanchuk [b,c], M. El Marssi [b]

[a] IMED-Lab, Cadi Ayyad University, Marrakesh 40000, Morocco
[b] LPMC, University of Picardy Jules Verne, Amiens 80039, France
[c] Physics Faculty, Southern Federal University, Rostov-on-Don 344090, Russia
[d] Laboratoire des Sciences et Métiers de l'Ingénieur (LSMI), Département Matériaux et Procédés, ENSAM Meknès, Université Moulay Ismail, Meknès, Morocco

ABSTRACT

Barium sodium niobate Ba$_2$NaNb$_5$O$_{15}$ (BNN) ceramics with different amounts of BaO-Na$_2$O-Nb$_2$O$_5$-P$_2$O$_5$ (BNNP) glass were prepared via the conventional solid-state method. The effect of glass content on structural, microstructure and dielectric properties of BNN ceramics was investigated. The XRD results showed that no secondary phase was formed after adding BNNP glass. It was found that such additions reduce the average grains size and refine the microstructure of the obtained ceramics. Moreover, the samples exhibited a stable dielectric constant over the temperature range of 25 °C–150 °C and their dielectric constants were significantly improved. The ceramic with 7.5 wt% BNNP glass content showed a dielectric constant which is more than twice as much as that of pure BNN ceramic, as well as a low dielectric loss less than 5%.
© 2021 Elsevier Ltd. All rights reserved.
Selection and peer-review under responsibility of the scientific committee of the International Conference on Phosphates (ICP): Fundamentals, Processes and Technologies.

## 1. Introduction

Nowadays, high energy storage density capacitors used as renewable and clean energy sources are attracting substantial attentions and studies [1]. Electrical ceramics have been extensively applied in the electronics and electrical industry, for electron beam, nuclear technique, hybrid electric vehicles (HEV), and other energy storage devices [2-4]. The most key energy storage components of pulse power techniques are ceramic capacitors [5,6]. Among these requirements, breakdown strength (BDS) is the primary barrier to limit their application in high-power and electronic devices [7,8] because the energy storage density depends not only on the dielectric permittivity ($\varepsilon_r$) but also on the electric breakdown strength of the material [6]. Within the various energy storage materials, dielectric ceramics exhibit relatively high dielectric constant and low BDS. Thus, the most key energy storage components of pulse power techniques are ceramic capacitors with high breakdown strength and high dielectric permittivity [9,10], making them promising candidates for energy-storage devices that can be even used in extremely working conditions [11].

In this context, the environmentally friendly Ba$_2$NaNb$_5$O$_{15}$ (BNN) ceramics are known ferroelectric materials that present a relatively high saturation polarization Ps (40 μC/cm$^2$) [12,13]. However, they exhibit relatively low breakdown strength due to the existence of defects, such as pores and impurities. Moreover, BNN exhibits a complicated sequence of structural phase transitions, the following sequence is normally accepted [14]: P4bm (T1 = 160 °C), Pba2 (T2 = 270 °C), P4bm (T3 = 560 °C), P4/mbm. To avoid high energy consumption, several approaches have been investigated in order to synthesize such materials at lower sintering temperatures; these have included the addition of sintering aids [15,16], low melting point oxides [17] and doping of other cations in solid solution [16]. Some studies have used galss additives [9,18]. However, it is important to select suitable additives to enhance the breakdown strength of ceramics without sacrificing high dielectric constant. In this context, Jun Song et al. [11] found that the addition of crystallizable BaO–SrO–Nb$_2$O$_5$–Al$_2$O$_3$–B$_2$O$_3$–SiO$_2$ glass could significantly lower the sintering temperature, refine the grain size and enhance the breakdown strength of Ba$_{0.39}$Sr$_{0.61}$Nb$_2$O$_6$ ceramics. Likewise, Su et al [19] reported that the crystallizable glass additive (BaO-TiO$_2$-SiO$_2$-Al$_2$O$_3$) could maintain a high dielectric constant (~2300) for the Ba$_{0.4}$Sr$_{0.6}$TiO$_3$ (BST) ceramics due to the precipitation of barium titanate from the glass phase during the sintering process, and could significantly improve the breakdown strength of BST ceramics. In addition, Wu et al. [20] have investigated the addition of glass–ceramic BaO-TiO$_2$-SiO$_2$ to Ba$_{0.4}$Sr$_{0.6}$TiO$_3$ ceramics. The findings showed that the crystal phase inhibited the decrease of the dielectric constant of ceramics. It appears from the above literature that the addition of crystallizable glasses to bulk ceramics can significantly improve energy storage properties, as well as solve the problem of reducing the dielectric constant.

* Corresponding author.
E-mail address: ihyadn.abderrahim@gmail.com (A. Ihyadn).

In our previous work [21], we found that the BaO–Na$_2$O–Nb$_2$O$_5$–P$_2$O$_5$ (BNNP) glass can precipitate Ba$_2$NaNb$_5$O$_{15}$ phase after crystallization heat-treatment. Thus, in this paper we choose BNNP glass as a sintering additive crystallizable glass and study its effect on structural, microstructure and dielectric properties of Ba$_2$NaNb$_5$O$_{15}$ (BNN) ceramic.

## 2. Experimental procedure

The Ba$_2$NaNb$_5$O$_{15}$ powders used in this study were prepared by the conventional solid-state synthesis route. Analytical reagent grade BaCO$_3$, Na$_2$CO$_3$, and Nb$_2$O$_5$ were used as starting materials. These materials were weighed according to the stoichiometric formula, and powder mixtures were ball mixed for 4 h in a high density polyethylene bottle with ethanol. Afterwards, the slurry was dried and the mixture was calcined in an alumina crucible at 1200 °C for 10 h in air and followed by second ball milling. Meanwhile, BaO-Na$_2$O-Nb$_2$O$_5$-P$_2$O$_5$ glass powders were prepared according to our previous work [21]. The two powders were weighed according to the nominal composition of (1-x)BNN-xBNNP and mixed in ethanol, then the obtained final powder mixtures were pressed into disks with a diameter of 13 mm. Final sintering was carried out at 1250 °C for 15 min and then dropped to 1000 °C for 6 h. An X-ray diffractometer (Panalytical™ X-Pert Pro spectrometer) with CuKα radiation (λ = 1.5405 Å) in a wide range of Bragg's angle (2θ) (10° ≤ 2θ ≤ 80°) at a scanning rate of 2°/min, was used to characterize the structure of the (1-x)BNN-xBNNP ceramics, the BNNP glass and the BNNP glass-ceramics at room temperature. The microstructure of the (1-x)BNN-xBNNP ceramics was observed using a scanning electron microscopy (SEM, Tescan VEGA3). For the impedance and electrical measurements, the sintered samples were painted a silver paste and fired at 500 °C for 20 min to form the electrode. The dielectric measurements were carried out at temperatures ranging from 25 to 400 °C with a rate of 3 °C/min in the heating and cooling process by using an impedance analyzer (LCR meter hp 4284A 20 Hz-1 MHz).

## 3. Results and discussion

### 3.1. X-ray diffractions

Fig. 1 (a) shows the XRD patterns of the BNN ceramics specimens with different glass additions. It can be found that a stable single phase with a tetragonal tungsten bronze structure (JCPDS#40-1463) can be formed in the (1-x)BNN-xBNNP ceramics and no any secondary phase is detected. The tetragonal tungsten bronze structure has been discussed by Sambasiv Rao [22] and by Run Li et al. [23]. Fig. 1(b) presents the XRD patterns of both BNNP glass and BNNP crystallized at 1000 °C for 6 h. As we can see from this figure, the XRD pattern of glass BNNP consists of large halo at low diffraction angles, which confirm the amorphous nature of the glass. However, the XRD pattern of the BNNP glass annealed at 1000 °C for 6 h reveals two main phases: Ba$_2$NaNb$_5$O$_{15}$ with tungsten bronze structure and NaNbO$_3$ with perovskite structure in accordance with our previous work [21].

### 3.2. Microstructures

Fig. 2 shows the SEM surface morphologies of Ba$_2$NaNb$_5$O$_{15}$ ceramics with different BNNP glass additives sintered at 1000 °C for 6 h. From Fig. 2 (a), the SEM image of BNN ceramic shows a significant amount of columnar grains and a low degree of porosity. Compared with pure BNN, when the glass addition increases from 0 to 7.5 wt%, the samples exhibit a dense microstructure with smaller and less pores. However, when the glass content is higher than 7.5 wt%, larger pores appeared, the grain boundaries became unclear and some melting phenomenon is observed in Fig. 2 (e). On the other hand, it is to be noted that glass additions have resulted in the decrease of the average grain size. The average grain size of (1-x)BNN-xBNNP ceramics decreases from 2.7 μm to 1.3 μm, 2.2 μm, 2.4 μm, 1.7 μm with an increase of × from 0 wt% to 2.5 wt%, 5 wt%, 7.5 wt%, and 10 wt% respectively. This could be attributed to the appearance of a liquid phase during sintering, which results in the inhibition of grain growth for BNN ceramics [9].

### 3.3. Dielectric properties

The temperature dependence of the dielectric constant for BNN ceramics at fixed frequencies of 0.5, 1, 5, 10, 100 kHz and 1 MHz over the temperature range from room temperature to 400 °C are shown in Fig. 3(a). The dielectric anomaly detected at temperatures close to 260 °C was correlated to one phase transition [24-26], which associated with the tetragonal (Bbm2) to orthorhombic (P4bm) phase transformation. A diffuse maximum, which was detectable especially at low frequency. Such a maximum is not observed the measurements carried out at at higher frequencies (>10 kHz). This phenomenon is consistent with the frequency-dependent dielectric constant that increases as the frequency decreases. Dielectric relaxation peaks can be associated with short-distance jumping of oxygen vacancies, since oxygen vacancies are active at low frequencies and act as "polarons", similarly to dipole reorientation [27,28]. Fig. 3 (b) presents the temperature dependence of dielectric constant on heating and cooling run on 3 °C/min at 1 kHz. The dielectric anomaly was identified at 255–233 °C during the heating and cooling cycles, respectively. The phase transition was thermally hysteresis with a ΔT of 22 °C. The same result was observed by ABell et al [29].

Fig. 4 (a) gives the temperature-dependent dielectric constant of BNN ceramics with different glass content measured at 1 kHz on heating process. It is obvious that all the curves exhibit one dielectric anomaly at around 260 °C. Dielectric constant is strongly influenced by the glass addition and increase gradually with glass content. For instance, the dielectric constant of pure ceramic is ~136, and samples containing 2.5, 5, 7.5 and 10 wt% glass content have a dielectric constant of 163, 236, 350 and 179 respectively. The introduction of a crystallizable glass BNNP phase could be responsible for this increase. The temperature dependence of dielectric loss (tan δ) of BNN ceramics with various glass content is shown in Fig. 4 (b). It is observed that, the dielectric loss is less dependent on the glass content, and the (1-x)BNN-xBNNP ceramics have relatively lower dielectric loss values (<0.05) in the temperature range of 25–100 °C. The enhanced dielectric constant and the low dielectric losses obtained could improve the energy storage properties of the materials.

The frequency dependence of the dielectric constant of BNN ceramic with various glass content measured at room temperature in the range of 100 Hz–1 MHz is shown in Fig. 5 (a). It is evident that the dielectric constants for all samples exhibit quite frequency stability in the whole measured frequency range.

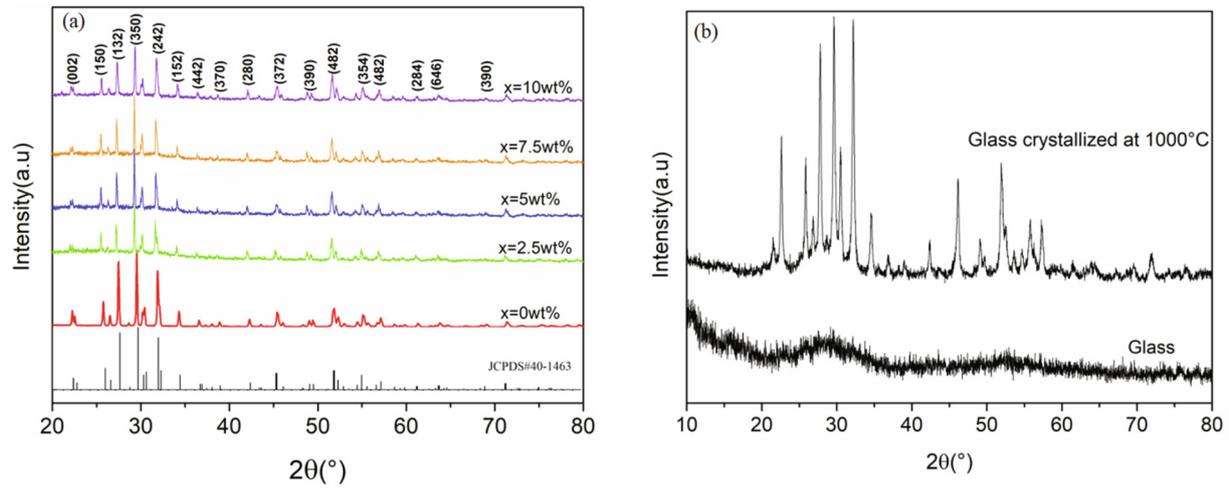

**Fig.1.** (a) XRD patterns of BNN ceramics with different glass compositions of BNNP glass, (b) XRD patterns of both BNNP glass and BNNP crystallized at 1000 °C for 6 h.

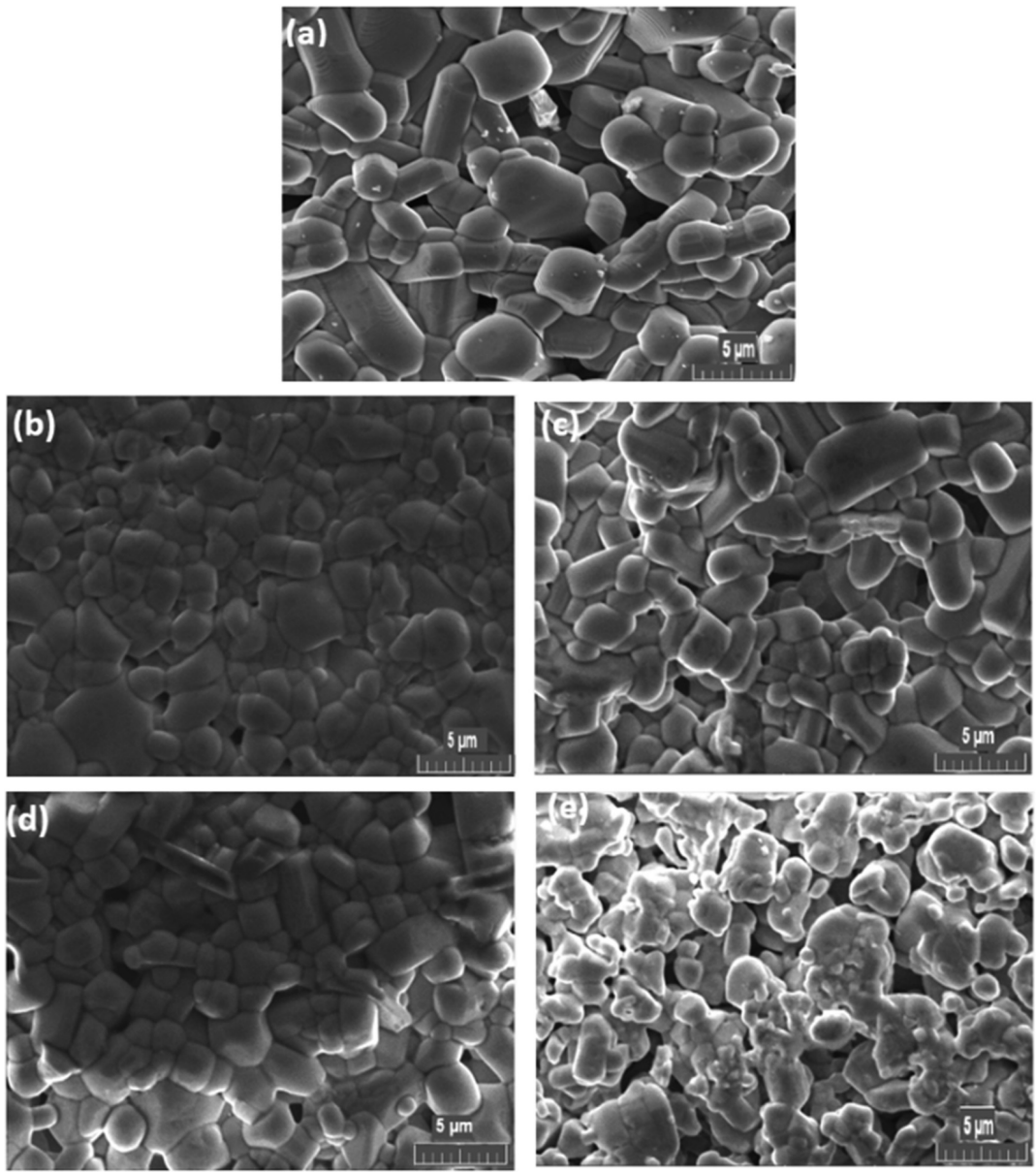

**Fig. 2.** SEM images of polished and thermally etched BNN ceramics with different glass content; (a) x = 0, (b) x = 2.5 wt%, (c) x = 5 wt%, (d) x = 7.5 wt%, (e) x = 10 wt%.

A. Ihyadn, D. Mezzane, M. Amjoud et al.

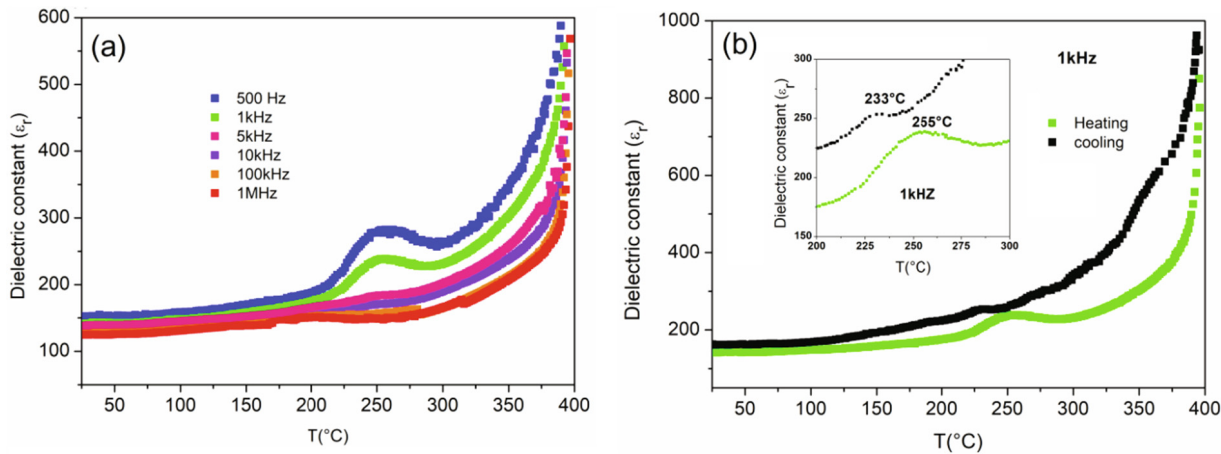

**Fig. 3.** (a) Temperature dependence of the dielectric constant of BNN ceramic at different frequencies on heating process, (b) temperature dependence of dielectric constant on heating and cooling run on 3 °C/min at 1 kHz.

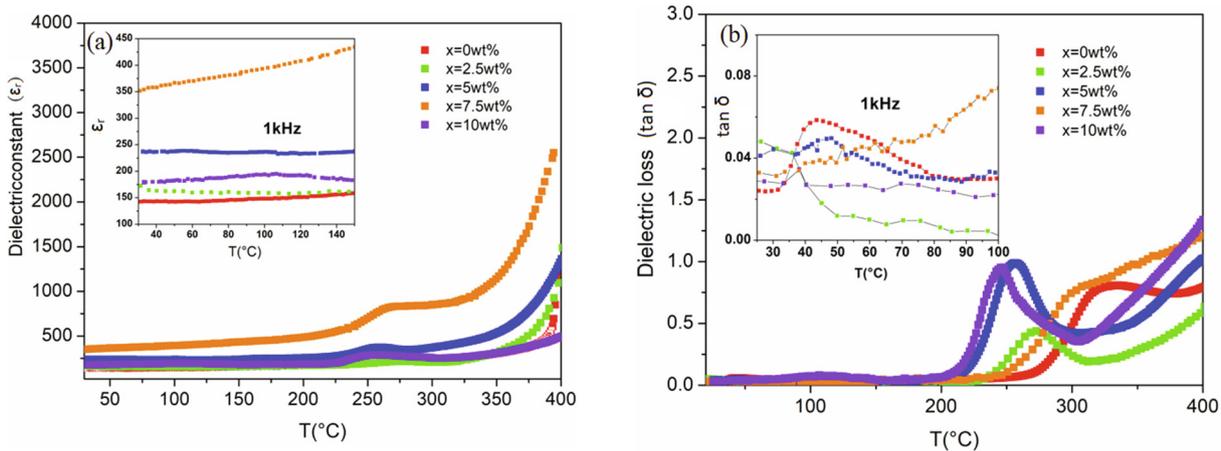

**Fig. 4.** (a) Temperature dependence of (a) dielectric constant and (b) dielectric loss of BNN ceramic with different glass content measure at 1 kHz and on heating process.

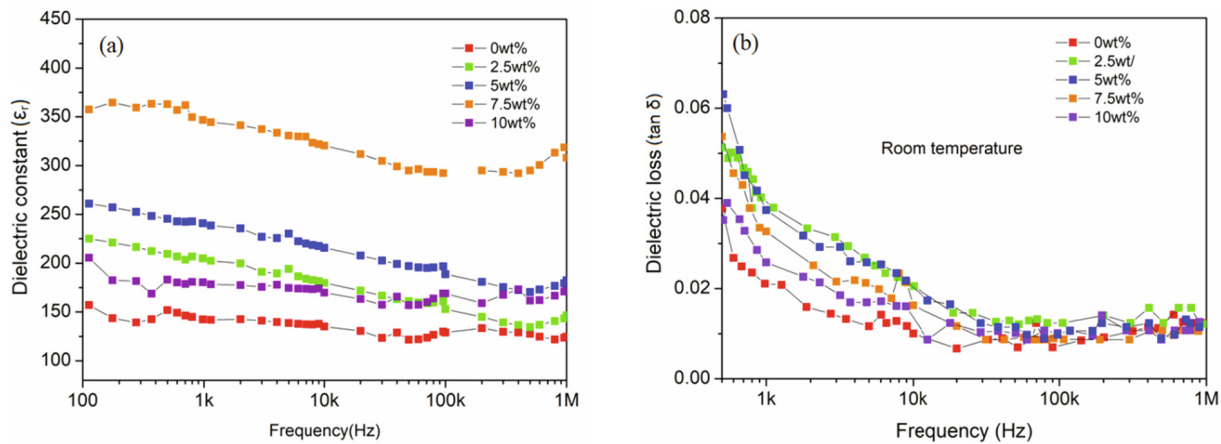

**Fig. 5.** (a) Frequency dependence of dielectric constant and (b) dielectric loss of BNN ceramics with different glass content measured at room temperature.

The frequency dependence of the dielectric loss of BNN ceramics with various glass content measured at room temperature in the range of 500 Hz–1 MHz is shown in Fig. 5(b). The dielectric losses tend to be stable in the frequency range from 10 kHz to 1 MHz after decreasing slightly from 500 Hz to 10 kHz. However, the dielectric loss values keep a very low level (<0.025) in the range of 10 kHz–1 MHz. These results indicate that the addition of BNNP glass significantly improved the dielectric properties of BNN ceramics in agreement with their microstructures.

## 4. Conclusion

$Ba_2NaNb_5O_{15}$ (BNN) ceramics with different $BaO$-$Na_2O$-$Nb_2O_5$-$P_2O_5$ (BNNP) glass content were prepared via solid state reaction method. X-ray diffraction analysis of these BNN ceramics shows the formation of tetragonal tungsten bronze structures without any detectable secondary phase in their structures. It is evidenced their microstructures that the glass addition leads decrease both the porosity and the grain size. It is found that the dielectric constant of the BNN ceramic with 7.5 wt% glass was remarkably enhanced and is almost twice higher than that of the pure BNN sample.

## CRediT authorship contribution statement

**A. Ihyadn:** Investigation, Writing - original draft, Visualization. **D. Mezzane:** Conceptualization, Validation, Resources, Supervision. **M. Amjoud:** Writing - review & editing, Investigation. **A. Lahmar:** Writing - review & editing. **L. Bih:** Conceptualization, Resources, Supervision. **A. Alimoussa:** Software, Supervision. **I.A. Luk'yanchuk:** Writing - review & editing. **M. El Marssi:** Formal analysis, Resources.

## Declaration of Competing Interest

The authors declare that they have no known competing financial interests or personal relationships that could have appeared to influence the work reported in this paper.

## Acknowledgments

The authors gratefully acknowledge the financial support of CNRST, OCP foundation and the European Union's Horizon H2020-MSCA-RISE research and innovation actions,-ENGIMA and MELON.